\documentclass[fleqn,twoside]{article}

\usepackage{espcrc2}

\readRCS
$Id: espcrc2.tex,v 1.2 2004/02/24 11:22:11 spepping Exp $
\usepackage{graphicx}
\usepackage[figuresright]{rotating}

\newcommand{\AmS}{{\protect\the\textfont2
  A\kern-.1667em\lower.5ex\hbox{M}\kern-.125emS}}

\title{Hints of new physics in bottomonium decays and spectroscopy
\thanks{Research under grant FPA2002-00612 and GV-GRUPOS03/094.}}
\author{Miguel Angel Sanchis-Lozano\thanks{Email:Miguel.Angel.Sanchis@uv.es}
\vspace{0.4cm}\\
Instituto de F\'{\i}sica
Corpuscular (IFIC) and Departamento de F\'{\i}sica Te\'orica,
Centro Mixto Universidad de Valencia-CSIC \\
Dr. Moliner 50, E-46100 Burjassot, Valencia (Spain)}

\begin{document} 
\begin{abstract}
A non-standard light
CP-odd Higgs boson could induce a slight (but observable) lepton 
universality breakdown in Upsilon leptonic decays. Moreover,  
the mixing between such a pseudoscalar Higgs boson and  $\eta_b$ states might
shift the mass levels of the latter, thereby changing the
values of the $m_{\Upsilon(nS)}-m_{\eta_b(nS)}$ splittings predicted 
in the standard model. Besides, also the $\eta_b$ width could  
be broader than expected, with potentially 
negative consequences for its discovery
in both $e^+e^-$ and hadron colliders.
        
\end{abstract}
\maketitle

\section{Lepton symmetry breaking}

In many extensions of the standard model (SM), new scalar 
and pseudoscalar states 
appear in the physical spectrum. Admittedly, the masses of these particles 
are typically of the same order as the weak scale and, in principle, a 
fine-tunning is required to make them much lighter. Nevertheless, if the theory
possesses a global symmetry, its spontaneous breakdown gives rise to a
massless Goldstone boson, the $\lq\lq$axion''.  
The original axion was introduced in the framework
of a two-Higgs doublet model (2HDM) \cite{gunion} to solve the  
strong CP problem. However, such an axial U(1) symmetry is anomalous and the 
pseudoscalar acquires a (quite low) mass ruled out experimentally.

On the other hand, if the global symmetry is explicitly (but slightly) broken, 
one expects a pseudo-Nambu-Goldstone boson in the theory which, for a 
range of model parameters, still can be significantly lighter than the 
other scalars. A good example is 
the so-called next to minimal supersymmetric standard model (NMSSM) 
where a new singlet superfield is added to the Higgs sector \cite{gunion}.
The mass of the lightest CP-odd Higgs can 
be naturally small
due to a global symmetry of the Higgs potential only softly broken by
trilinear terms \cite{Hiller:2004ii}. Moreover, the 
smallness of the mass is protected
from renormalization group effects in the region of large $\tan{\beta}$ 
(defined as a ratio of two Higgs vacuum expectation values). 
Actually, there are other scenarios containing a light 
\footnote{By $\lq\lq$light'' we
consider here a broad interval which might reach a
${\cal O}(10)$ GeV mass value}
pseudoscalar Higgs boson which could have escaped detection in the searches
at LEP-II, e.g. a MSSM Higgs sector
with explicit CP violation \cite{Carena:2002bb}. Another example is 
a minimal composite Higgs scenario
\cite{Dobrescu:2000yn} where the lower bound on the CP-odd scalar mass
is quite loose, as low as $\sim 100$ MeV (from astrophysical constaints).

In this work we consider a possible New Physics (NP) contribution
to the leptonic decays of $\Upsilon$ resonances below
$B\bar{B}$ threshold via the decay modes: 
$$\Upsilon{\rightarrow}\ {\bf \gamma_s}\ 
\eta_b^* (\rightarrow 
A^0 {\rightarrow}\ \ell^+\ell^-)\ ;\ 
\ell=e,\mu,\tau
$$
where $\gamma_s$ stands for a 
soft (undetected) photon, $A^0$ denotes
a non-standard light CP-odd Higgs boson and
$\eta_b^*$ a spin-singlet
$b\bar{b}$ virtual state. Here we will mainly focus
on the $\Upsilon(1S)$ state.

Our later development is based upon the following keypoints: 
\begin{itemize}
\item Such a NP contribution would be
unwittingly ascribed to the 
leptonic branching fraction (BF) of Upsilon resonances
\item A leptonic (squared) mass dependence in the width
from the Higgs contribution
would lead to an $\lq\lq$apparent'' lepton universality breakdown
\end{itemize}

\begin{figure}
\begin{center}
\includegraphics[width=14pc]{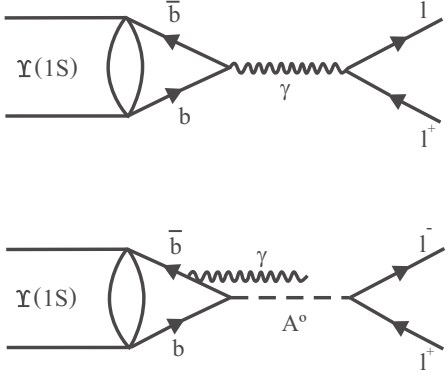}
\end{center}
\caption{
(a): Conventional electromagnetic annihilation of
the $\Upsilon(1S)$ resonance into 
a $\ell^+\ell^-$ pair. (b): Annihilation mediated by a 
CP-odd Higgs boson ($A^0$) subsequent 
to a M1 transition of the $\Upsilon$ into a 
virtual $\eta_b^*$ state.}
\end{figure}

Our theoretical analysis relies on the factorization
of the decay width:
\[
{\Gamma}_{\Upsilon \to \gamma_s\ell\ell}=
\frac{\Gamma^{M1}(\Upsilon \to \eta_b^*\gamma_s)
\times {\Gamma}_{\eta_b^*\to\ell\ell}}{\Gamma}
\]
where $\Gamma$ is an unknown parameter to be interpreted
as a width; 
${\Gamma}_{\eta_b^*\to\ell\ell}$ denotes the annihilation 
width of the intermediate $\eta_b^*$ state
into a lepton pair; the width of a M1 transition is
given in the nonrelativistic approximation by \cite{Sanchis-Lozano:2003ha}: 
$\Gamma^{M1}_{\Upsilon{\rightarrow}\gamma_s \eta_b^*} \simeq 
4\alpha Q_b^2k^3/{3m_b^2}$,  
with $Q_b$ and $m_b$ denoting the electric charge and mass of the
bottom quark respectively; 
$\alpha$ stands for the fine structure constant
and $k$ is the soft photon energy, approximately equal to
the (yet unknown) hyperfine splitting $m_{\Upsilon}-m_{\eta_b}$,
assumed to be in the range $\simeq 35-150$ MeV.

For $\Gamma$ close to $\Gamma_{\eta_b^*}$,
the ratio $\Gamma_{\eta_b^*\to\ell\ell}/\Gamma$
may be interpreted as the $\eta_b^*$ branching fraction
into a lepton pair; the width of the
whole process is 
\begin{equation}
\Gamma_{\,\Upsilon\to\gamma_s\,\ell\ell}\ =\ 
\Gamma^{\,M1}_{\,\Upsilon\to\gamma_s\eta_b^*}\ 
\times\ \frac{\Gamma_{\eta_b^*\to\ell\ell}}{\Gamma_{\eta_b^*}}
\label{eq:factor1}
\end{equation}
leading to the cascade decay formula
\[
BF[{\,\Upsilon\to\gamma_s\,\ell\ell}]\ =\ 
BF[{\,\Upsilon\to\gamma_s\eta_b^*}] \times
BF[\eta_b^* \to \ell \ell] 
\]
This result can be obtained using a
time-ordered perturbative calculation \cite{Sanchis-Lozano:2003ha}
for an almost on-shell intermediate $\eta_b^*$ state - consistent
with the emission of a soft photon.

On the other hand, higher Fock components beyond 
the heavy quark-antiquark pair can play
an important role in both production and decays 
of heavy quarkonium \cite{Bodwin:1994jh}. 
In fact, 
$\Gamma^{M1}_{\Upsilon{\rightarrow}\gamma_s \eta_b^*}/\Gamma$
may be interpreted as the  
probability ${\cal P}^{\Upsilon}(\eta_b^*\gamma_s)$
that a $\eta_b^*+\gamma_s$ configuration exists as a Fock state
inside the $\Upsilon$ resonance during a typical time of order $1/k$
much longer than the typical annihilation time, of order
$1/m_b$ \cite{Sanchis-Lozano:2003ha}. Thus, the $b\bar{b}$
annihilation would eventually free a quasi-real photon $\gamma_s$.

Therefore, we will 
factorize the decay width as
\begin{equation}
\Gamma_{\,\Upsilon\to\gamma_s\,\ell\ell}=
{\cal P}^{\Upsilon}(\eta_b^*\gamma_s) \times \Gamma_{\eta_b^*\to\ell\ell}
\label{eq:factor2}
\end{equation}
in accordance with a non relativistic effective theory
\cite{Bodwin:1994jh}. Note that the processes underlying
factorizations of Eqs.~(\ref{eq:factor1}) and
\ref{eq:factor2} should be competitive.

\begin{table*}[hbt]
\setlength{\tabcolsep}{0.4pc}
\caption{Measured leptonic branching fractions 
${\cal B}_{\ell\ell}$ and error bars (in $\%$) of
$\Upsilon(1S)$ and $\Upsilon(2S)$ 
(from \cite{pdg}).}

\label{FACTORES}

\begin{center}
\begin{tabular}{ccccc}
\hline
channel: & $e^+e^-$ & $\mu^+\mu^-$ & $\tau^+\tau^-$ &  ${\cal R}_{\tau}$ \\
\hline
$\Upsilon(1S)$ & $2.38 \pm 0.11$ & $2.48 \pm 0.06$ & $2.67 \pm 0.16$ & 
$0.10 \pm 0.07$\\
\hline
$\Upsilon(2S)$ & $1.34 \pm 0.20$ & $1.31 \pm 0.21$ & $1.7 \pm 1.6$ & 
$0.28 \pm 1.21$ \\
\hline
\end{tabular}
\end{center}
\end{table*}

\section{Estimates according to a 2HDM(II)}

In order to make numerical estimates we
will assume that fermions couple to the $A^0$ field
according to the effective Lagrangian
\[
{\cal L}_{int}^{\bar{f}f}\ =\ -\xi_f^{A^0}\ \frac{A^0}{v}
m_f\bar{f}(i\gamma_5)f 
\]
with $v \simeq 246$ GeV and $\xi_f^{A^0}$ depends on
the fermion type. In this work we focus on a  
2HDM of type II \cite{gunion}, whence $\xi_f^{A^0}=\tan{\beta}$ for down-type
fermions; thus
\cite{Sanchis-Lozano:2002pm,Sanchis-Lozano:2003ha}
\[
{\Gamma}_{\eta_b^*\to\ell\ell}\ =\ 
\frac{3}{32\pi^2Q_b^2\alpha^2}
\frac{m_b^4m_{\ell}^2\tan{}^4\beta}{{(1+2x_{\ell})\Delta}m^2v^4}\ {\times}\ 
\Gamma_{\ell\ell}^{(em)}
\]
where $\Delta m = | m_{A^0}-m_{\eta_b} |$ and the
electromagnetic decay width into a dilepton is given 
by the Van-Royen Weisskopf formula: 
\[
{\Gamma}^{(em)}_{{\ell}{\ell}}= 
4\alpha^2Q_b^2\ \frac{|R_n(0)|^2}{M_{\Upsilon}^2}\ {\times}\ 
K(x_{\ell})
\]
where $K(x_{\ell})=(1+2x_{\ell})(1-4x_{\ell})^{1/2}$
is a (smoothly) decreasing function of 
$x_{\ell}=m_{\ell}^2/M_{\Upsilon}^2$ with $m_{\ell}$ the lepton mass.
Only for the tauonic mode would the NP contribution to the $\Upsilon$
leptonic decay be significant.

\begin{figure}
\begin{center}
\includegraphics[width=15pc]{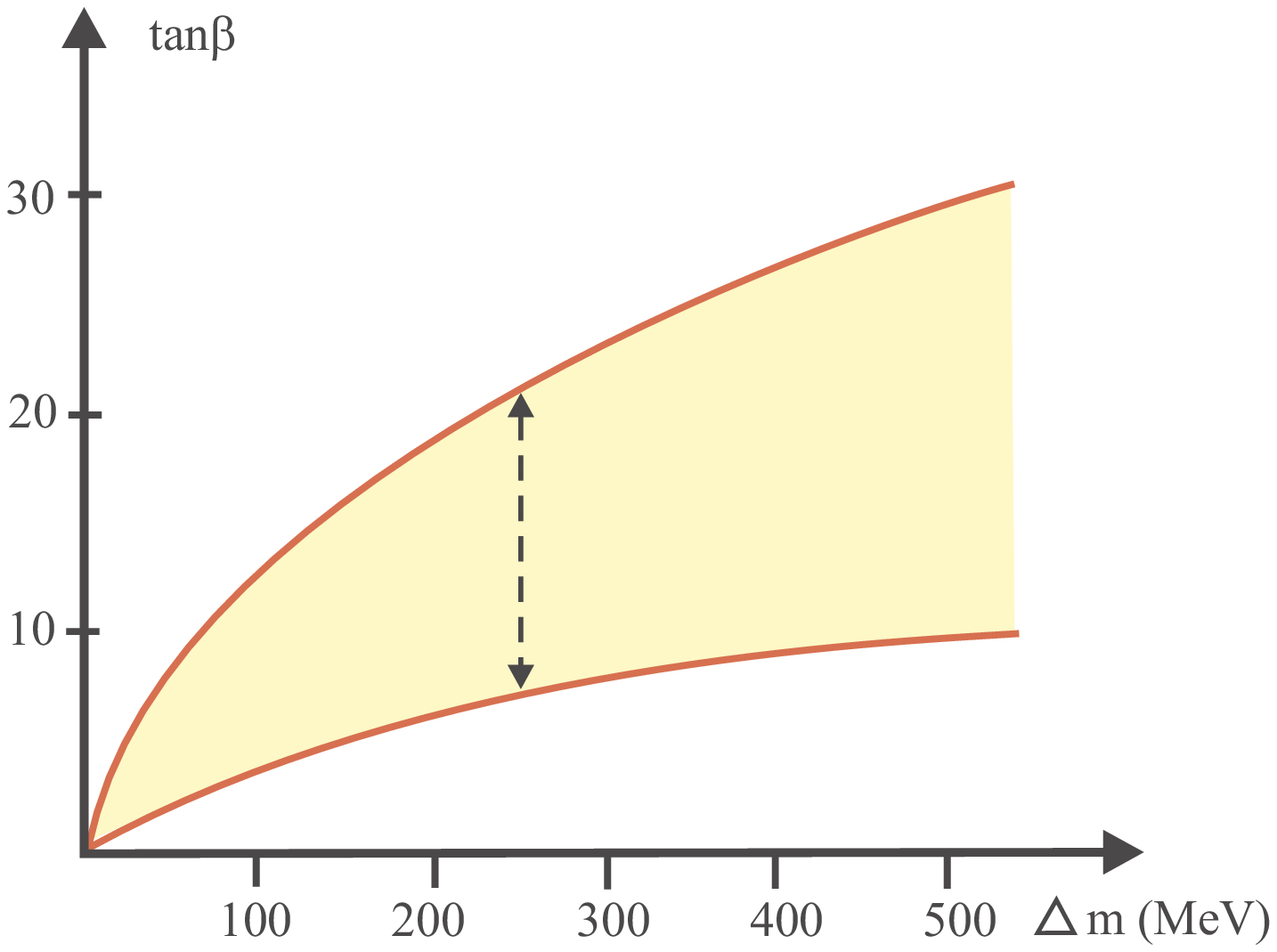}\\
\includegraphics[width=15pc]{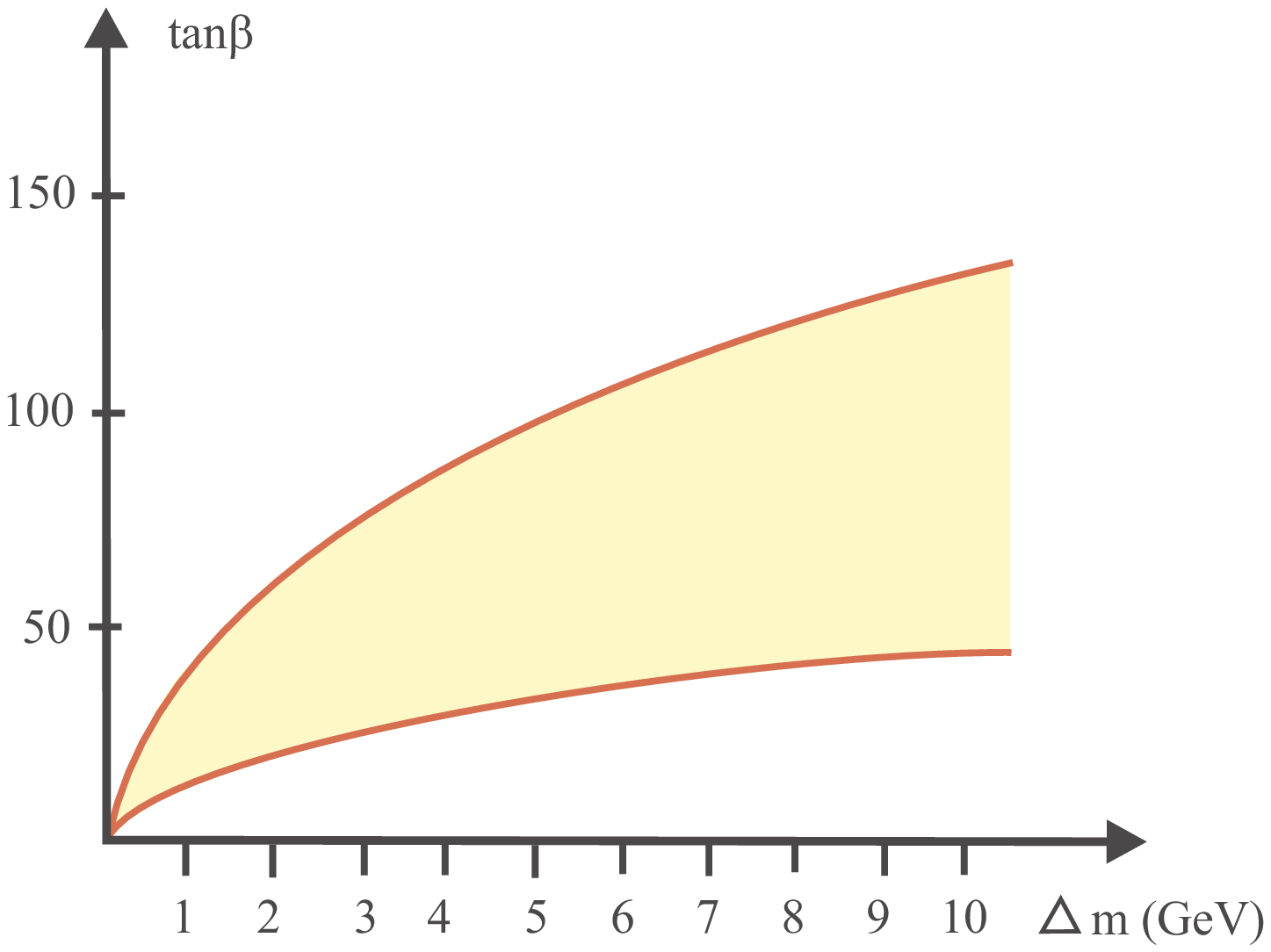}
\end{center}
\caption{$\tan{\beta}-\Delta m$ values (shaded area) required to
yield a ${\cal O}(10)\%$ lepton universality breaking effect using
the factorization Eq.~(\ref{eq:factor2})  and
setting $\Gamma=50$ keV \cite{Sanchis-Lozano:2004zd}. 
The arrow in the upper plot indicates
the range of $\tan{\beta}$ using $\Delta m=250$ MeV 
\cite{Sanchis-Lozano:2003ha}. The upper and lower curves in both plots
correspond to the minimal ($k=35$ MeV) and maximal ($k=150$ MeV) estimates 
of the  M1-transition probability ${\cal P}^{\Upsilon}(\eta_b^*\gamma_s)$, 
respectively.
For (relatively) large values of $\Delta m$ (i.e. large Higgs mass values)
only the lower values of the shaded region are acceptable,
corresponding to the highest estimates of  
${\cal P}^{\Upsilon}(\eta_b^*\gamma_s)$.}
\end{figure}

To check our conjecture we define the ratio:
\[
{\cal R}_{\tau}=\frac{\Gamma_{\Upsilon{\rightarrow}\gamma_s\tau\tau}}
{\Gamma^{(em)}_{\ell\ell}}=
\frac{{\cal B}_{\tau\tau}-\bar{\cal B}_{\ell\ell}}
{\bar{\cal B}_{\ell\ell}}
\]
where $\bar{\cal B}_{\ell\ell}=({\cal B}_{ee}+{\cal B}_{\mu\mu})/2$
stands for the mean BF of the electronic and muonic modes. A
(statistically significant) non-null value of 
${\cal R}_{\tau}$ would imply the rejection
of lepton universality (predicting ${\cal R}_{\tau}=0$) and a strong argument
supporting the existence of a pseudoscalar Higgs boson.

If the factorization of Eq.~(\ref{eq:factor1}) is adopted assuming 
$\Gamma \simeq \Gamma_{\eta_b^*} \simeq \Gamma_{\eta_b}$, one gets
\[ \Gamma_{\,\Upsilon\to\gamma_s\,\tau\tau}\ =\ 
\Gamma^{\,M1}_{\,\Upsilon\to\gamma_s\eta_b^*}\ 
\times\ \frac{\Gamma_{\eta_b^*\to\tau\tau}}{\Gamma_{\eta_b^*}}
\]
For large $\tan{\beta}$ ($\geq 35$) the
NP contribution would almost saturate the $\eta_b^*$ decay:
$\Gamma_{\eta_b^*} \simeq \Gamma_{\eta_b^*\to\tau\tau}$; thus
\[ {\cal R}_{\tau} \simeq
\frac{\Gamma^{\,M1}_{\,\Upsilon\to\gamma_s\eta_b^*}/
\Gamma_{\Upsilon}}{\bar{\cal B}_{\ell\ell}}
\simeq 1-10\ \% \]
for $k=50-150$ MeV.

Instead relying on the factorization of Eq.~(\ref{eq:factor2}), one gets 

\[ {\cal R}_{\tau} \simeq \biggl[\frac{m_b^2k^3\tan{}^4\beta}
{8\pi^2{\alpha}(1+2x_{\tau})\Gamma_{\Upsilon} v^4}\biggr] 
\times \frac{m_{\tau}^2}{{\Delta}m^2}
\]

Current experimental data (see Table 1) indicate that 
there might be 
a difference of order $10\%$ in the BF's
between the tauonic channel on the one side, and the electronic and muonic
modes on the other side. The range 
of $\tan{\beta}$ 
needed to account for such an effect, applying 
the factorization of Eq.~(\ref{eq:factor2}), is shown in Fig.~2 as a function
of the mass difference ($\Delta m$) between the postulated 
non-standard Higgs boson and the $\eta_b(1S)$ resonance. 
For the factorization of
Eq.~(\ref{eq:factor1}) with $\tan{\beta} \geq 35$, an agreement can be found
for $k > 50$ MeV.

\section{Possible spectroscopic consequences}

The mixing of the $A^0$ with a pseudoscalar resonance 
could modify the properties of both 
\cite{Drees:1989du,opal}. In particular, it might cause
a disagreement between the 
experimental determination of
the $m_{\Upsilon(nS)}-m_{\eta_b(nS)}$ hyperfine splittings  
and theoretical predictions based on quark potential
models, lattice NRQCD or pQCD. 
The masses of the mixed (physical) states in terms of
the unmixed ones (denoted as $A_0^0, \eta_{b0}$) are:
\begin{eqnarray}
m_{\eta_b,A^0}^2 & \simeq & \frac{1}{2}(m_{A_0^0}^2
+m_{\eta_{b0}}^2) \nonumber \\
& \pm & \frac{1}{2}\biggr[\ (m_{A_0^0}^2
-m_{\eta_{b0}}^2)^2+4(\delta m^2)^2\ \biggl]^{1/2} \nonumber
\end{eqnarray}
where $\delta m^2  \simeq  0.146 \times \tan{\beta}$ GeV$^2$
\cite{Drees:1989du}.
For some mass intervals, the above formula simplifies to:
\begin{eqnarray}
m_{\eta_b,A^0} & \simeq & 
m_{\eta_{b0}}\ \mp\ \frac{\delta m^2}{2m_{\eta_{b0}}}\ ; \nonumber \\
&& 0 < m_{A_0^0}^2-m_{\eta_{b0}}^2 << 2\ \delta m^2, \nonumber \\
m_{\eta_b,A^0} & \simeq & 
m_{\eta_{b0}}\ \mp\ \frac{(\delta m^2)^2}
{2m_{\eta_{b0}}(m_{A_0^0}^2-m_{\eta_{b0}}^2)}\ ; \nonumber \\
&&  m_{A_0^0}^2-m_{\eta_{b0}}^2 >> 2\ \delta m^2 \nonumber
\end{eqnarray}
Setting $\tan{\beta}=20$ and  
$m_{\eta_{b0}} \simeq m_{A_0^0}=9.4$ GeV, as an illustrative example, one gets
$m_{\eta_b} \simeq 9.24$ GeV and
$m_{A^0} \simeq 9.56$ GeV yielding 
$BF[\Upsilon(1S) \to \gamma \eta_b(1S) \simeq 10^{-2}$. A caveat is 
thus in order: a 
quite large $m_{\Upsilon}-m_{\eta_b}$ difference
may lead to an unrealistic ${\cal P}^{\Upsilon}(\eta_b^*\gamma_s)$,
requiring smaller $\tan{\beta}$ values, in turn 
inconsistently implying a smaller
mass shift; hence no hyperfine splitting 
greater than $\sim 200$ MeV should be expected.

\section{Summary}
In this paper, possible
hints of new physics in bottomonium systems
have been pointed out:

a) Current experimental data do not preclude the possibility of 
lepton universality breaking 
at a significance level of $10\%$, 
interpreted in terms of a light CP-odd Higgs boson for 
a reasonable range of $\tan{\beta}$ values

b) Mixing between the CP-odd Higgs and $\eta_b$ states can 
yield $m_{\Upsilon(nS)}-m_{\eta_b(nS)}$ splittings larger 
than expected within the SM if $m_{A_0^0} > m_{\eta_{b0}}$;
the opposite if $m_{A_0^0} < m_{\eta_{b0}}$

c) Broad $\eta_b$ widths are also 
expected for high $\tan{\beta}$ values. All
that might explain the failure to find any signal from 
hindered $\Upsilon(2S)$ and $\Upsilon(3S)$ magnetic dipole transitions
into $\eta_b$ states. There could be also
negative effects on the prospects to
detect $\eta_b$ resonances in hadron colliders like the Tevatron
through the decay modes:
$\eta_b \to J/\psi+J/\psi$ \cite{Braaten:2000cm},
and the recently proposed $\eta_b \to D^*D^{(*)}$ 
\cite{Maltoni:2004hv}, 
as the respective BF's would drop by about one order of magnitude
with respect to the SM calculations

d) New results on tauonic BF's of all three $\Upsilon(1S)$,
$\Upsilon(2S)$, $\Upsilon(3S)$ from CLEO on-going analysis are 
eagerly awaited

\end{document}